# Nature of carrier injection in metal/2D semiconductor interface and its implications to the limits of contact resistance

*Divya Somvanshi[1,⊥,∥], Sangeeth Kallatt[1,2,∥], Chenniappan Venkatesh[1], Smitha Nair[2], Garima Gupta[1], John Kiran Anthony[3], Debjani Karmakar[4], and Kausik Majumdar[1,\*]*

[1]Department of Electrical Communication Engineering, Indian Institute of Science, Bangalore 560012, India

[2]Center for Nano Science and Engineering, Indian Institute of Science, Bangalore 560012, India

3Renishaw India, Bangalore 560011, India

[4]Bhabha Atomic Research Center, Homi Bhabha National Institute, Trombay, Mumbai: 400085, India

[⊥]Currently with Department of Physics and Astronomy, Georgia State University, Atlanta, Georgia 30303, USA

[∥]Equal contribution

[*]Corresponding author, email: kausikm@iisc.ac.in

**ABSTRACT:** Monolayers of transition metal dichalcogenides (TMDCs) exhibit excellent electronic and optical properties. However, the performance of these two-dimensional (2D) devices are often limited by the large resistance offered by the metal contact interface. Till date, the carrier injection mechanism from metal to 2D TMDC layers remains unclear, with widely varying reports of Schottky barrier height (SBH) and contact resistance ($R_c$), particularly in the monolayer limit. In this work, we use a combination of theory and experiments in Au and Ni contacted monolayer $MoS_2$ device to conclude the following points: (i) the carriers are injected at the source contact through a cascade of two potential barriers – the barrier heights being determined by the degree of interaction between the metal and the TMDC layer; (ii) the conventional Richardson equation becomes invalid due to the multi-dimensional nature of the injection barriers, and using Bardeen-Tersoff theory, we derive the appropriate form of the Richardson equation that describes such composite barrier; (iii) we propose a novel transfer length method (TLM) based SBH extraction methodology, to reliably extract SBH by eliminating any confounding effect of temperature dependent channel resistance variation; (iv) we derive the Landauer limit of the contact resistance achievable in such devices. A comparison of the limits with the experimentally achieved contact resistance reveals plenty of room for technological improvements.

# I. INTRODUCTION

Two-dimensional transition metal dichalcogenides (TMDCs) are promising materials for novel electronic and optoelectronic device applications (*1,2*). Monolayer and few-layers of these materials have been shown to be very promising for light generation (*3,4*) and detection (*5,6*) applications. On the other hand, the ability to suppress surface roughness scattering at sub-nm thickness, coupled with an appreciable bandgap, makes them promising candidate to enable logic transistor scaling beyond 10 nm (*7*). However, the performance of most of these electronic and optoelectronic devices are bottlenecked by a relatively large parasitic contact resistance (*8,9,10,11*). On the contrary, in photodetection applications, metal/TMDC interface plays active role in enhancing photoresponse (*12,13*). Thus, it is important to understand the nature of the interface between the metal and the TMDC in these devices. However, despite its importance, there has been a limited effort to understand the origin of the intrinsic mechanisms that control the characteristics of such interface.

The Schottky barrier height and the contact resistance of a metal/TMDC interface have been reported in the past (*14*) (*15,16,17,18,19,20,21,22*) (*23,24,25*), however, with a large spread, particularly at the monolayer limit (*8,26,27,28,29,30*). In this work, we use a combination of ab-initio theory, systematic experiments, and modeling to reveal the underlying mechanisms that control the SBH and the contact resistance. With the help of ab-initio calculation and material characterization, we first study the modification of electronic properties of monolayer $MoS_2$ underneath the metal contact due to metal induced charge transfer. Next, we propose a two-barrier carrier injection model arising from such charge transfer between contact metal and 2D material underneath. We then derive the corresponding modified Richardson equation of such composite barrier based on Bardeen-Tersoff theory. The magnitude of the effective potential barrier is obtained experimentally by using a novel TLM based extraction methodology, and is found to be a strong function of device operating condition. This extraction method carefully excludes any ambiguity resulting from channel resistance variation due to temperature. Further insights into the mechanism is obtained by validating the experimental results with solution of 1-D coupled Poisson-Schrodinger equations. Finally, the Landauer limit (*35*) of the contact resistance achievable in such structure is derived analytically and compared with the experimental results. All the results described below are based on Au and Ni contacted monolayer $MoS_2$ devices, but the conclusions remain qualitatively valid for a generic top contacted layered semiconductor interface if the thickness of the semiconductor remains close to the two-dimensional limit.

## II. NATURE OF CARRIER INJECTION AT SOURCE JUNCTION

Figure 1 shows the schematic diagram of a monolayer $MoS_2$ film on $SiO_2$/Si substrate and contacted by metal pad from top. Along the channel direction, such a contact, in general, has been treated like a conventional metal-semiconductor band bending (*14*) with strong Fermi level pinning and tunneling induced field emission. However, note that, there is no provision for band bending vertically downward in the ultra-thin sandwich layer. This forces the carrier injection mechanism as a cascade of multiple processes (*30,36*), as schematically shown in Figure 1: (i) vertically downward injection of carriers to the thin film underneath the metal, where the 2-D film is modified electronically due to proximity of metal, (ii) horizontal transport of carriers through the modified 2-D film underneath the metal contact (the current crowding regime), and (iii) horizontal injection of carriers to the channel over a second barrier arising from the doping difference between the 2-D films under the metal contact and the 2-D film in the channel. Consequently, such a cascaded carrier injection process can be modeled as a series combination of two diodes and a resistor at the source end. Before going into the details of the carrier injection mechanism through such cascaded processes, we first discuss the properties of the monolayer MoS$_2$ film underneath the metal contact.

## III. MODIFIED ELECTRONIC PROPERTIES OF MONOLAYER MOS$_2$ IN PROXIMITY OF METAL

Clearly, the 2-D film underneath the contact plays the mediating role between the source and the channel in the carrier injection process. Former ab-initio calculations have been performed to study interfaces between TMDC monolayers and metal surfaces (*31,32,33,34*). We use ab-initio calculation to obtain the insights into the charge transfer between metal (Au or Ni) and monolayer MoS$_2$. As sulfur vacancies (SV) are considered as one of the most probable defects in MoS$_2$, we consider both with and without SV cases. This is followed by optical and electrical characterization of monolayer MoS$_2$ film in proximity of metal.

### A. Ab-initio calculation details

The MoS$_2$/Au interface is constructed by combining the 4×4×1 monolayer of Au [111] surface on 4×4×1 monolayer of 1H- MoS$_2$. Due to the mismatch in the lattice parameters of Au [111] and 1H- MoS$_2$, there is an initial lateral strain of ~ 9.7% on the Au layer, which, after structural relaxation, turns out to be 6.08%. In a similar procedure, keeping a minimal lattice parameter incongruity, MoS$_2$/Ni interface is built by combining a 5×5×1 monolayer of Ni [111] surface on 4×4×1 monolayer of 1H- MoS$_2$ with 1.5% initial lateral strain lateral strain on Ni-layer, which, after structural relaxation remains almost the same. To avoid replication from the *z*-directional periodicity, a vacuum of 10 Å was added both above and below the constructed interface. These interfaces are also investigated in presence of ~ 6% sulfur vacancies (SV), known to be the most common defect to occur in 1H- MoS$_2$. Thus, we investigate four such interfaces, *viz*. Au, Au + SV,

Ni and Ni + SV with 1H- MoS$_2$. To resemble the realistic experimental scenario, we have also investigated the interfaces of monolayer MoS2 and bilayer metals.

These interfaces were explored with the help of ab-initio Density-functional theory (DFT) based formalism using plane-wave pseudopotential approach using projector augmented wave (PAW) potentials as implemented in Vienna Ab-initio Simulation Package (VASP). Electron correlation within the system is treated using Perdew-Burke-Ernzerhoff exchange-correlation functional under spin-polarized generalized-gradient approximation (GGA). Ionic and lattice parameter optimization of the constructed interfaces are obtained by conjugate gradient algorithm until the Hellmann-Feynman forces on each ion is less than 0.01eV/Å. To account for the interface-induced dipolar interaction, we have incorporated the van der Waal corrections by using Grimme DFT-D2 method (*37*). In this method, a semi-empirical dispersion potential is added to the conventional density functional energy after taking care of inter-surface interactions. For self-consistent calculations and structure optimization, an energy cut-off of 500 eV is used with a *k*-point mesh size 5×5×3.

### B. Ab-initio results

The layer (LPDOS) and orbital (OPDOS) projected density of states are summarized in Figure 2(a)-(e), and the corresponding charge densities are shown in Figure 2(f)-(i). Covalent charge sharing between Ni and S is more favourable than Au, as Au has relatively more closed shell structure, while Ni has a partially filled valence 3*d*-orbital. We have started with a distance of 3.2 Å between both of the layers for all four cases. After relaxation, the average distance between Au and 1H- MoS$_2$ becomes ~ 3 Å. The value for Ni turns out to be ~ 2.2 Å. The average distance reduces to 2.8 Å and 2.0 Å for Au + SV and Ni + SV cases respectively, implying proximity of the metal layer to 1H-MoS$_2$ in presence of SV.

In monolayer MoS$_2$, the π-bonded S-$p_x$, $p_y$ and Mo-$d_{xy}$, $d_{x2-y2}$ and $d_{3z2-1}$ orbitals populate the states at the top of the valence band and bottom of the conduction band in the bonding and antibonding manifold respectively. Au layer, having a filled 5$d$ orbital and delocalized 6$s$ electrons, transfers its 6$s$ electrons to the Mo-4$d$ orbitals via S-3$p_x$ and 3$p_y$. A close observation of Figure 2(b) unveils that the Au-S hybridized bonding orbitals populate energy states from -1.5 to -3 eV, which are filled Au-5$d$ states. The charge transfer from Au-6$s$ to S or Mo renders a delocalization of Au-6$s$ electrons. The states from -1.5 eV to -0.4 eV are having more Mo-4$d$ and S-3$p$ bonding characters than Au-6$s$, implying the already occurred charge transfer from Au layer to MoS$_2$. These transferred electrons fill up the Mo-$d_{xy}$, $d_{x2-y2}$ and $d_{3z2-1}$ orbitals and thereby shifts the $E_F$ towards the conduction band. Due to such complete charge transfer, probability of formation of gold sulphide is less. In presence of SV, the shift of $E_F$ is more, as the absence of S-3$p$ orbital allows direct charge transfer from Au-6$s$ to Mo-4$d$, implying an increase in the extent of $n$-type doping. Bader analysis for charge transfer has revealed that for Au and Au+SV cases, there is an average charge transfer of 0.8e and 1.5e per atom respectively from Au to the MoS$_2$ layer. This interfacial charge in calculated after integrating the charge over a volume around the interface between two specific z-values. Less amount of charge at the Au/MoS$_2$ interface (than Ni/MoS2 interface as discussed later) and more localized charge distribution at the individual layer suggests tunnelling nature of charge transfer for Au/MoS$_2$ contacts (*38*).

The presence of partially filled 3$d$ orbitals makes Ni a better contact for MoS$_2$. Ni/MoS$_2$ interface, both with and without SV, exhibit metallic property [Figure 2(d)-(e)] with a strong Fermi level shift towards the conduction band, due to the covalent nature of the charge transfer from Ni-3$d$ to S-3$p$ and formation of nearly compensated antiferromagnetic metal nickel sulphides at the interface. In contrary to Au, for Ni, S-3$p$ DOS is highly delocalized and hybridizes with Ni-3$d$ OPDOS. Amount of charge transfer from Ni to MoS$_2$ layer is 0.5e and 1.2e for Ni and Ni+SV cases respectively. The interfacial charge for Ni interface is ~ 4 times higher than the Au-interface. This result can be visualized from the charge density plots presented in Figure 2(h) and (i), indicating much higher charge overlap and bonding of Ni and S ions.

As a next-step, we have investigated the impact of increasing the thickness of metal-layer by constructing interfaces of MoS$_2$ monolayer and metal bilayer. The OPDOS for these systems are depicted in Figure 3(a) and (b). The corresponding charge density plots are presented in Figure 3(c) and (d) respectively. For MoS$_2$/Au interface, increment of metal-layer thickness leads to more delocalized Au-6$s$ states in an energy range of -1 to +1 eV around $E_F$, mainly having contributions from the top Au-layer. Delocalized electrons from the layer adjacent to MoS$_2$ have already transferred the *n*-type carriers to the beneath MoS$_2$ layer. Due to the non-spin polarized nature of 6$s$ electrons, the system remains non-magnetic irrespective of increase of Au-layer thickness, as is evident from Figure 3(a). Interface with Ni, on the contrary, have strong impact on the spin-polarization of the system. The Ni-layer adjacent to MoS$_2$ forms antiferromagnetic Nickel sulphide at the interface, leading to the partial DOS of Ni-3$d$ electrons almost same at $E_F$ for both spin-up and spin-down states. With increasing thickness of Ni-layer, the ferromagnetic nature of bulk-Ni prevails over the interfacial antiferromagnetic nature, resulting into more spin-polarization at $E_F$. This trend becomes obvious from Figure 3(b). The spin-polarized Ni-3$d$ states near $E_F$ is less hybridized with S-3$p$ states, implying less covalence of additional Ni-layer with S. Nature of charge transfer is similar to the monolayer metal cases, as can be seen from Figure 3(c) and (d), since the adjacent layer contributes the most in charge transfer. Interfacial charge remains more for Ni-contact than Au.

For the sake of completeness, we intend to study whether the nature of metal contacts remains same with increasing thickness of $MoS_2$. Therefore, in Supplemental Material S1, we discuss the DFT results of a bilayer metal/bilayer $MoS_2$ system and observe that the above-mentioned metal/$MoS_2$ interfacial effects remain qualitatively similar.

### C. Experimental results – optical characterization of charge transfer

To support the analysis of metal induced charge transfer effects on electronic properties of monolayer $MoS_2$, we now experimentally characterize a monolayer thick $MoS_2$ film in close proximity of metal. However, in a typical "$MoS_2$-bottom/metal-top" contact structure, it is difficult to characterize the inaccessible $MoS_2$ film underneath the metal. To avoid this problem, we prepare two sets (Au and Ni) of "metal-bottom/$MoS_2$-top" structures, as mentioned below. In this structure, we characterize the $MoS_2$ film both optically, as well as using KPFM, while maintaining the proximity of metal states.

To obtain the proposed structure, periodic structures of 4 μm wide Au, and separated by 4 μm are obtained on a 285-nm thick $SiO_2$ layer on Si substrate using photolithography, followed by e-beam evaporation of metal and subsequent lift-off. A similar sample is prepared for Ni as well. Monolayers of $MoS_2$ layers are exfoliated on top of this, and only those monolayers are selected which connect at least two parallel metal lines. The thickness of the $MoS_2$ flake is confirmed by optical contrast in a microscope on the $SiO_2$ portion and also by measuring the separation between the $A_{1g}$ and the $E^1_{2g}$ Raman peaks.

Photoluminescence (PL) and Raman spectra were taken using a 532-nm laser focused using 100X objective. The Raman shift of the $A_{1g}$ peak for monolayer $MoS_2$ on metals [Figure 4(a)-(b)] shows a larger broadening compared with a sample on $SiO_2$, while the $E^1_{2g}$ peak broadening remains almost substrate independent. Such broadening can be attributed to anharmonicity due to laser induced heating (*39*) and substrate induced doping (*40*). However, larger broadening for monolayer samples on metals cannot attributed to the heating effect owing to better heat conduction by metal compared with $SiO_2$. This suggests that the additional broadening occurs due to metal induced charge transfer effect. By comparing the obtained Raman peak shift and full width at half maximum (FWHM) with the data presented in ref. (*40*), we estimate the doping density in our monolayer $MoS_2$ samples to be $\approx 2.5 \times 10^{12}$ cm$^{-2}$ and $\approx 9.7 \times 10^{11}$ cm$^{-2}$ for Ni and Au substrates, respectively. Also, the photoluminescence intensity of $A_{1s}$ exciton peak is found to be dramatically suppressed for monolayers on metals [Figure 4(c)], supporting interlayer charge transfer between metal and $MoS_2$.

We perform XPS analysis of MoS2 on $SiO_2$ and metal substrates, and the estimated Fermi level shift agrees qualitatively with the DFT predicted shifts (Supplemental Material S2). The XPS data indicates that the binding energies of the different core levels of $MoS_2$ on the metal substrates are blue shifted compared to the $SiO_2$ substrate samples. This is an indication of a relative shift of the Fermi level closer to the conduction band edge for the samples on Au and Ni substrates, suggesting *n*-type doping.

### D. Effect of charge transfer – creation of cascaded potential barriers

Owing to such metal induced charge transfer, as depicted in Figure 1, we introduce a model for electron injection from the source through three cascaded processes: (i) overcoming a vertical thermionic barrier height $\phi_V$ coupled with a tunneling barrier width $d$ due to the vdw gap between the top metal contact and the *modified* MoS$_2$ underneath the contact [diode D$_V$, shown in Figure 5(a)]; (ii) horizontal transport through MoS$_2$ underneath the contact resulting in a current crowding resistance $R_d$ [Figure 5(b)]; and finally (iii) overcoming a horizontal barrier $\phi_H$ [diode D$_H$, shown in Figure 5(c)] between *modified* MoS$_2$ under the contact and the undoped MoS$_2$ in the channel. Based on our ab-initio calculation, two different scenarios may occur in determining $\phi_H$: a metal contact, like Au, introduces limited gap states into the monolayer under the contact, and $\phi_H$ originates primarily due to doping difference (between MoS$_2$ under contact and MoS$_2$ channel) induced built-in potential [left panel of Figure 5(c)]. On the other hand, a metal contact, like Ni, induces a large density of gap states, which in turn results in strong Fermi level pinning at the MoS$_2$ under contact and channel MoS$_2$ junction, and hence a Schottky barrier [right panel of Figure 5(c)].

Note that, apart from charge transfer induced doping, the vertical barrier $\phi_V$ is also modulated due to the image force experienced by a carrier at the MoS$_2$ layer underneath the contact (*41,42*) owing to the close proximity of conducting metal layer. For our top contact structure, as shown in Supplemental Material S3, we estimate a barrier height lowering ($\Delta\phi_V$) of 0.25 eV due to image force.

To understand the nature of the different potential barriers, the Kelvin Probe Force Microscopy (KPFM) images and the corresponding Atomic Force Microscopy (AFM) images of the samples are shown in Figure 4(d)-(e) for Au [and Figure 4(g)-(h) for Ni]. All the metal lines were grounded during KPFM measurement. The KPFM results indicate the contact potential difference (CPD) between the tip and the sample, which allows us to infer the local work function differences $\Delta W = W_{tip} - W_{sample}$, where $W_{tip} = 5.3$ eV. Using scan line 1 [Figure 4(d)-(f)], we note that $W_{tip} - W_{Au} \approx 0$, hence $W_{Au} = 5.3$ eV. We also see that $W_{Au} - W_{MoS_2/Au} = 0.12$ eV, inferring $W_{MoS_2/Au} = 5.18$ eV. Using scan line 2, we obtain $W_{MoS_2/SiO_2} - W_{MoS_2/Au} = 0.13$ eV, which directly implies that the SBH between the two regions ($\phi_H$) is 0.13 eV. For the Ni sample [Figure 4(g)-(i)], we have similarly obtained $W_{Ni} = 5.06$ eV, $W_{MoS_2/Ni} = 5.18$ eV, and $\phi_H = 20$ meV. As Raman data suggests that Ni causes higher doping than Au, a lower $\phi_H$ for Ni suggests a formation of Schottky barrier in $D_H$ due to large density of bandgap states in MoS$_2$ under Ni. Heavier source doping caused by Ni would have otherwise resulted in larger $\phi_H$ in a purely doping difference induced barrier.

## IV. CARRIER INJECTION MODEL AT THE SOURCE FOR 2D/METAL CONTACT – A MODIFIED RICHARDSON EQUATION:

**Need for a modification in Richardson equation:** Richardson equation is often used for characterization of metal-semiconductor Schottky barriers and extraction of the corresponding barrier height (*41*). However, the form of the Richardson equation to be used must conform with the dimensionality of the contact structure, as explained in Figure 6. For example, when a 3-D metal is contacted with a bulk 3-D semiconductor (Figure 6a), the degree of the polynomial factor in $T$ is 2 (*41*), which reduces to 1.5 when a 2-D semiconductor is edge contacted with a 3-D metal (*43*), as shown in Figure 6b. However, in our present case of a top contacted 2-D semiconductor by a 3-D metal (Figure 6c), the carrier injection occurs through three cascaded processes, as mentioned before. A corresponding Richardson equation for such a structure is missing. Hence, we derive below a carrier transport model for each of the three processes, and then combine them in cascade to obtain an effective Richardson equation.

**Vertical contact diode:** To obtain the effective Richardson equation that describes the vertical charge injection mechanism, we take the MoS$_2$ portion underneath the metal contact as a doped ultra-thin semiconductor. We model D$_V$ [Figure 5(a)] as two different planes of carriers are separated by a vdw gap $d$. In the *Appendix,* we derive a modified Richardson equation for the vertical diode D$_V$ by following the approach of Bardeen (*44*) and Tersoff (*45*):

$$J_v = A'_V e^{-2k_0 d} T^{\alpha_V} e^{-q\phi_V/k_B T} = A^*_V T^{\alpha_V} e^{-q\phi_V/k_B T} \qquad (1)$$

with $\alpha_V = 1$ and $\phi_V = \phi_{V0} - \eta V_g - \gamma V_{ds}$. $\phi_{V0}$ is the barrier height without any bias, $V_g$ is the back gate voltage, $V_{ds}$ is the applied drain bias grounding the source, $\eta$ and $\gamma$ are screening dependent parameters (see Appendix). $A^*_V = A'_V e^{-2k_0 d}$ is the modified Richardson constant and $k_0$ is a constant defined in the Appendix.

**Current crowding regime:** We understand the origin of current crowding effect and a corresponding resistance $R_d$ as follows: As shown in the inset of Figure 5(b), due of continuity of current flow, at any point, the loss due to the vertical current must be compensated by reduction in the horizontal current. Thus, at i$^{th}$ node, we obtain the voltage $V(x)$ from the continuity equation: $\sigma \left[ \frac{V_{i-1} - V_i}{\Delta x} - \frac{V_i - V_{i+1}}{\Delta x} \right] = J_v(x) \Delta x$ where $\sigma$ is the in-plane conductivity of the 2D material under the metal. This reduces to $\frac{d^2 V(x)}{dx^2} = \frac{\zeta}{\sigma} e^{qV(x)/k_B T}$ with $\zeta = A^*_V T^{\alpha_V} e^{-q(\phi_{V0} - \eta V_g)/k_B T}$. $V(x)$ can be solved analytically to obtain

$$V(x) = -\frac{2}{\beta} \ln \left[ \cos \left\{ \frac{\beta}{2} (D_0 + \lambda x) \right\} \right] \qquad (2)$$

where $\beta = \frac{q}{k_B T}$, $\lambda = \left(\frac{2\zeta}{\beta\sigma}\right)^{1/2}$, and $D_0 = -\frac{2}{\beta}\tan^{-1}\left(\sqrt{e^{\beta V_0} - 1}\right)$. Here, $q$ is the magnitude of electron charge, $k_B$ is the Boltzmann constant, $T$ is the temperature, and $V_0 = V(x = 0)$. $V(x)$ underneath the contact is plotted in Figure 5(b) showing the extent of current transfer.

**Horizontal diode:** We model $D_H$ [Figure 5(c)] as a two-dimensional diode in the plane of the monolayer MoS$_2$ with a barrier height $\phi_H$. The corresponding Richardson equation is obtained by using the approach described in ref. (*43*):

$$J_H = A_H^* T^{\alpha_H} e^{-q\phi_H/k_B T} \tag{3}$$

with $\alpha_H = 1.5$.

**Combined effect – Effective Richardson equation:** Note that, both $\alpha_V$ and $\alpha_H$ differ from the conventional value of $\alpha = 2$ for three-dimension (*41*). The combined processes are schematically depicted in Figure 5(d). The injected electrons experience an effective barrier due to cascading effect of $D_V$, $R_d$ and $D_H$, following an effective Richardson equation:

$$J = A_{eff}^* T^{\alpha_{eff}} e^{-q\phi_{B,eff}/k_B T} \tag{4}$$

where the effective barrier height $\phi_{B,eff} \leq \phi_V + \phi_H$. The equality holds only when either of $\phi_V$ and $\phi_H$ is negligible (that is, zero reflection from one of the diodes) and the transport underneath the metal is near ballistic. Also, since $\phi_V$ and $\phi_H$ vary with biasing conditions, $\alpha_{eff}$ varies in the range $1 \leq \alpha_{eff} \leq 1.5$.

# V. A TLM BASED METHODOLOGY FOR EFFECTIVE BARRIER HEIGHT EXTRACTION

**Existing issues in barrier height extraction:** In TMDC literature, both $\alpha_{eff} = 2$ (*16,17*) or $\alpha_{eff} = 1.5$ (*15,46*) have been used for barrier height extraction. Apart from the choice of $\alpha_{eff}$, the extracted SBH is usually confounded due to another reason: the channel resistance does not remain constant with temperature owing to strong dependence of carrier mobility on temperature in these thin layers (*29*). Hence Richardson equation does not completely describe the temperature dependence of the total current. The extracted barrier height is usually underestimated as mobility degrades with temperature.

**Proposed methodology:** To nullify the temperature dependent channel resistance effect, we fabricate a set of back gated and top metal contacted devices with varying channel length and use transfer length method (TLM). Monolayer $MoS_2$ flakes are first exfoliated on a Si wafer covered by 285 nm $SiO_2$, which acts as the back-gate dielectric. To have uniform rectangular channel, the monolayer flakes are first patterned by electron beam lithography followed by reactive ion etching (RIE) for 20 s in $BCl_3$ (15 sccm) and Ar (60 sccm), with an RF power of 50 W and chamber pressure of 4.5 mTorr, at $-10°$ C. Second level of electron beam lithography is used to define top metal contacts. Respective metals [namely, (i) 50 nm Au or (ii) 7 nm Ni / 50 nm Au) are deposited using electron beam evaporation at a $4.5 \times 10^{-7}$ Torr, followed by metal lift off in acetone. For the back gate, Aluminium is evaporated on the back side after a dilute HF treatment.

A set of devices with Raman mapping and SEM micrograph are shown in Figure 7(a). The $I_{ds}$-$V_g$ characteristics of a typical device with Au and Ni contacts are shown in Figure 7(b), at different temperatures (215 K to 290 K). The electrical measurements of the devices were done at a vacuum level of 2.25× $10^{-6}$ Torr using an Agilent B1500 device analyzer. All our Ni contacted devices exhibit lower threshold voltage compared with the Au contacted devices. This suggests strong Fermi level pinning close to the conduction band edge at the contact MoS$_2$/channel MoS$_2$ junction in turn electrically dopes the channel n-type (12). Figure 7(c) shows almost hysteresis-free drivability of 52 µA/µm in a 300-nm channel length Au/monolayer device at $V_g$ = 80 V and $V_{ds}$ = 2 V.

Figure 8(a) shows the method of extraction of the contact resistance, where the total contact resistance ($R_{cT}$) is obtained from a linear fit of total resistance: $R_T W = R_{sh} L + R_{cT} W$, and subsequent extrapolation of the line at $L = 0$. Here, $R_{sh}$ and $W$ are the sheet resistance of the channel and width of the devices, respectively. The extracted $R_{cT}$ is a strong function of device operating condition $(V_g, V_{ds})$. Note that, $R_{cT} = R_{cS} + R_{cD}$ is the total contact resistance offered by the source and drain sides, which are generally unequal. This is due to the presence of finite thermal barrier in the source side, while the electrons do not experience any such barrier on the drain side. However, at large $V_g$ and $V_{ds}$, the source side barrier is diminished, and both sides contribute almost equally to $R_{cT}$. Owing to lack of thermal barrier, the drain side is also expected to be weakly dependent on $V_g$ and $V_{ds}$, leading to an approximate estimation of the drain side contact resistance as $R_{cD} \approx 0.5 \times R_{cT}(V_{g,max}, V_{ds,max})$ which is bias independent. The source side component of $R_{cT}$ is thus extracted as $R_{cS}(V_g, V_{ds}) = R_{cT}(V_g, V_{ds}) - R_{cD} \approx R_{cT}(V_g, V_{ds}) - 0.5 \times R_{cT}(V_{g,max}, V_{ds,max})$. Thus, a current $I_c = \frac{V_{ds}}{R_{cS}}$ will be delivered by the source contact diode if the complete bias $V_{ds}$ is hypothetically applied across only the source contact diode. This is schematically explained in Figure 8(b). $I_c$ characterizes the source contact diode without any confounding effect from the channel resistance or the drain. Using $\alpha_{eff} = 1$, Figure 8(c) shows good linear fit between $\ln(I_c T^{-\alpha_{eff}})$ with $\frac{q}{k_B T}$, allowing us to unambiguously extract $\phi_{B,eff}$. The extracted $\phi_{B,eff}$ for Au is plotted as a function of $V_g$ and $V_{ds}$ in Figure 9(a) and (b). In Supplemental Material S4, the extraction is performed with $\alpha_{eff} = 1.5$ showing a difference of ~10 meV in the extracted barrier height. The Ni contacted devices [Figure 9(c) and (d)] consistently show lower $\phi_{B,eff}$ compared with Au, in agreement with the KPFM analysis discussed before.

**Validation with simulation:** To get insights into the $V_g$ dependence of $\phi_{B,eff}$, we solve 1-D coupled Poisson-Schrodinger (CPS) equations along the vertical direction of the device, both at the channel regime and at the contact regime (see Supplemental Material S5). From the simulated potential, the barrier height is extracted in Figure 10, which will be valid for small $V_{ds}$ due to 1D nature of the equations. Using only channel doping as a fitting parameter, we can obtain good agreement between the simulation (red lines) and the TLM extracted SBH. The deviation of $\phi_{B,eff}(V_g)$ from linearity at higher gate voltage arises due to strong gate field screening. Using KPFM extracted $\phi_H$ in Figure 4, and assuming $\phi_{B,eff} \approx \phi_V + \phi_H$, the individual components $\phi_V$ and $\phi_H$ are extracted at $V_g$=0. Also, using this $\phi_V(V_g = 0)$, we extract the doping of the monolayer MoS$_2$ film underneath the contact, and using the same, the $\phi_V$ at non-zero $V_g$ is also simulated (blue lines). At larger gate field, $\phi_H$ almost collapses, and $\phi_V$ dominates the total barrier. With an increase in $V_{ds}$, $\phi_V$ also is suppressed, and the device enters a *"zero thermal barrier"* regime of operation [bottom right corner of Figure 9(b) and (d)], where only $d$ and $R_d$ control the contact resistance.

# VI. CONTACT RESISTANCE AND ITS LANDAUER LIMIT FOR 2D TMDC/METAL CONTACT INTERFACE

In Figure 11, $R_{cT}$ of the Au and Ni devices is plotted as a function of the 2D sheet carrier density ($n_s$) at different temperatures. $n_s$ is obtained as $n_s = C_{ox}(V_g - V_t)/q$. At T=290 K, gate field = 0.167 V/nm and $V_{ds}$ = 1 V, the contact resistance ($R_c = R_{cT}/2$) between monolayer MoS$_2$ and Ni (Au) has been found to be 14 kΩ μm (23 kΩ μm), which reduces to 9 kΩ μm (for Ni) after vacuum anneal at 127°C for one hour. The extracted values of $R_{cT}$ from another set of Au and Ni TLMs are shown in Supplemental Material S6.

To investigate how these values compare with the fundamental limit of contact resistance achievable in these structures, we use Landauer approach (47) to find the conductance in a monolayer MoS$_2$/metal contact: $G = \left(\frac{2q^2}{h}\right) g_v M T_{eff}$, where $h$ is Plank's constant, $g_v$ is valley degeneracy (= 2 for monolayer MoS$_2$ arising from degenerate $K$ and $K'$ valleys), $M$ is number of current carrying modes per valley. $T_{eff}$ is the effective transmission probability through the three cascaded processes described earlier and is given by $T_{eff} = \frac{1}{\frac{1}{T_V}+\frac{1}{T_d}+\frac{1}{T_H}-2}$. For maximum transmission limit, $T_{eff,max} = e^{-2k_0 d}$ is obtained at the zero thermal barrier and ballistic limit, by noting that $T_{V,max} = e^{-2k_0 d}$ (tunneling through zero thermal barrier vdw gap in D$_V$), $T_{d,max} = 1$ (ballistic limit of MoS$_2$ underneath metal), and $T_{H,max} = 1$ (zero thermal barrier in D$_H$). At low temperature limit, only electrons at Fermi level take part in conduction, and hence $k_0 = \frac{(2m^*\varphi_{avg})^{1/2}}{\hbar}$ (See *Appendix*). Here we assumed $\varphi_{avg} = 0.5(W_m + \chi_{MoS_2})$. The number of modes of a two-dimensional conducting channel is given by (47): $M = Int\left[\sqrt{\frac{2m^* W^2 \Delta E}{\pi^2 \hbar^2}}\right] \approx W\sqrt{\frac{2n_s}{g_v \pi}}$ where $n_s$ is given by $n_s = g_v \frac{m^*}{\pi \hbar^2} \Delta E$ ignoring Fermi-Dirac broadening. Here $Int[.]$ is the maximum integer function. Hence,

$$R_{c,min}W = \frac{W}{G_{max}} = \left(\frac{h}{2q^2}\right) e^{2k_0 d} \sqrt{\frac{\pi}{2g_v n_s}}. \qquad (5)$$

In Figure 6(b), the $2R_{c,min}W$ limits are plotted for Au, Ni, and also for an ideal contact where $d = 0$ is assumed. For comparison, ideal multi-layer limit is also shown using $g_v = 6$. The obtained limits clearly allow provision for further technological improvement. Note that, at smaller $V_g$, the extracted $R_c$ decreases with an increase in temperature, due to enhanced thermionic emission efficiency. However, the trend diminishes and eventually reverses at higher bias (see Supplemental Material S7), where the thermionic emission efficiency does not change appreciably due to negligible barrier, but the carrier scattering under contact increases with an increase in temperature, in turn increasing contact resistance.

## VII. CONCLUSION

In conclusion, using a combination of theoretical and experimental techniques, we investigated the nature of the carrier injection at the junction between a monolayer $MoS_2$ and the contacting metal. We have shown that the charge transfer between contact metal and $MoS_2$ underneath plays a key role in such a contact, where the carrier from source is injected via two cascaded thermal barriers. The corresponding Richardson equation of such a Schottky diode requires appropriate modification in the power of T. At large gate and drain bias, both Au and Ni offer zero effective thermal barrier contact, where the contact resistance is limited by the tunneling vdw gap and the conductivity of the monolayer underneath the contact. At this zero-barrier condition, the fundamental lower limits of contact resistance are obtained theoretically using Landauer approach.

The insights obtained will be useful in designing well-behaved contacts for high performance two-dimensional electronic and optoelectronic devices.

## ACKNOWLEDGMENT

D. K. acknowledges support from BARC-ANUPAM supercomputing facility and help from Mr. Tuhin Kumar Maji. K.M. acknowledges the support of a start-up grant from IISc, Bangalore, the support of a grant under Space Technology Cell, ISRO-IISc, the support of grants under Ramanujan Fellowship, Early Career Award, and Nano Mission from the Department of Science and Technology (DST), Government of India.

## APPENDIX: DERIVATION OF RICHARDSON EQUATION FOR VERTICAL CURRENT INJECTION IN METAL-2D SEMICONDUCTOR VERTICAL JUNCTION (DIODE D$_V$)

The current injection mechanism in the vertical diode is a combined process of thermionic emission and tunneling through the vdw gap ($d$) between the metal surface and the 2D material. To model such carrier injection, we assume the 2D material as an almost perfect two-dimensional plane. Assuming effective mass approximation (free electrons in the plane of the metal surface and the 2D material), we write the wave function of the electron at the metal surface as

$$\psi_m = \Omega_m^{-1/2} e^{-k_m(d-z)} e^{i\overline{k_\parallel^m} \cdot \overline{\rho}} \quad (A1)$$

and at the 2D semiconductor as

$$\psi_s = \Omega_s^{-1/2} e^{-k_s z} e^{i\overline{k_\parallel^s}\cdot\bar{\rho}} \qquad (A2)$$

where z = 0 is assumed as the plane of the 2D semiconductor. $\Omega_m$ and $\Omega_s$ are respectively the metal and semiconductor area under consideration. Here $\bar{\rho}$ is the position vector in the plane of the metal surface or the 2D material, and $\overline{k_\parallel^m}$ ($\overline{k_\parallel^s}$) is the corresponding in-plane wave-vector in the metal (semiconductor) plane. $k_m$ and $k_s$ are the decay constant of the wave functions in the vdw gap between the metal and 2D material, and given by $k_m = \frac{[2m^*(W_m-\epsilon)]^{1/2}}{\hbar}$ and $k_s = \frac{[2m^*(\chi_s-\epsilon)]^{1/2}}{\hbar}$ at energy $\epsilon$. Here, $m^*$ is the carrier effective mass, $W_m$ represents the work function of the metal and $\chi_s$ is the electron affinity of the 2D semiconductor. The major contribution of the current comes from energy states close to the band edge of the 2D material, and hence $W_m, \chi_s \gg \epsilon$, allowing us to treat $k_m, k_s$ as independent of energy. Assuming elastic tunneling, and under small bias $V$, the tunneling current can be obtained using first order perturbation theory:

$$I = \frac{2\pi q}{\hbar}\sum_{ms}|C_{ms}|^2 f(E_m + qV)\{1 - f(E_s)\}\delta(E_m - E_s) \qquad (A3)$$

Here, $q$ is the magnitude of electron charge, $\hbar = h/2\pi$ with $h$ is Planck constant, $E_{m(s)}$ is the energy of corresponding state in the metal (2D semiconductor), $f(E)$ is the Fermi-Dirac probability of carrier occupation at energy $E$. The tunneling matrix element has been shown to be (*45*)

$$C_{ms} = -\frac{\hbar^2}{2m^*}\int d\bar{A}\cdot(\psi_m^*\bar{\nabla}\psi_s - \psi_s\bar{\nabla}\psi_m^*) \qquad (A4)$$

Using the wave functions in equations (A1) and (A2), one can evaluate the matrix element in equation (A4) to

$$C_{ms} = -\frac{\pi\hbar^2}{m^*\Omega}(k_m + k_s)e^{-k_m d}e^{(k_m-k_s)z}\ \delta(\overline{k_\parallel^m} - \overline{k_\parallel^s}) \qquad (A5)$$

where $\Omega = (\Omega_m \times \Omega_s)^{1/2}$. Using equations (A3) and (A5), and noting that only those electrons with an energy more than $q\phi_V$ can tunnel through the vdw gap, we can re-write the current density as

$$J_v = \kappa\frac{2\pi^3 q\hbar^3}{m^{*2}}(k_m + k_s)^2 e^{-2k_m d}e^{2(k_m-k_s)z} \sum_{\substack{k_\parallel^m,k_\parallel^s \\ E_m > q\phi_V}} [f(E_m + qV)\{1 - f(E_s)\}\delta(E_m - E_s)\delta(\overline{k_\parallel^m} - \overline{k_\parallel^s})] \qquad (A6)$$

where $\kappa$ is a normalization constant. The delta functions under the summation ensure conservation of in-plane momentum as well as conservation of energy during the elastic tunneling process. Note that, due to continuity equation, the current must be independent of $z$, which ensures that $k_m = k_s(=k_0)$. This is expected, as we are only considering elastic tunneling.

For further evaluation, we convert the summation over 2D k-space into integral over energy, as follows:

$$J_v = \kappa\frac{8\pi^3 q\hbar^3}{m^{*2}}k_0^2 e^{-2k_0 d} \int_{q(\phi_{V0}-\eta V_g)}^{\infty} dE\ D(E)e^{-(E+q\gamma V_{ds})/k_B T} \qquad (A7)$$

Here we approximated the Fermi-Dirac distribution as Boltzmann distribution, and also assumed $1 - f(E_s) \approx 1$. $k_B$ is the Boltzmann constant and $T$ is the temperature. The density of states is given by $D(E) = Min\{D_m(E), D_s(E)\} = D_s(E) = g_v \frac{m^*}{\pi\hbar^2}$ where $g_v$ is the valley degeneracy, and is 2 for monolayer MoS$_2$ arising from $K$ and $K'$ valleys. In Equation (A7), $V = \gamma V_{ds}$ is the difference between the quasi-Fermi levels of the metal and the portion of the 2D material under the metal, and depends on the applied bias $V_{ds}$. Similarly, application of a positive back gate voltage $V_g$ pushes down the conduction band edge of the 2D material underneath the metal, and hence suppresses the effective barrier height by $\eta V_g$. For small $V_g$, $\eta$ is a constant, and depends on how well the gate is electrostatically coupled to the 2D material under the metal contact. However, at larger $V_g$, $\eta$ reduces significantly owing to strong screening (*48*). $\phi_{V0}$ is the barrier height in the absence of any external bias. Also, assuming only states close to the Fermi level contribute to the current, we take $k_0^2 \approx \frac{2m^*\varphi_{avg}}{\hbar^2}$, where $\varphi_{avg} = 0.5(W_m + \chi_{MoS_2})$. Substituting in equation (A7), the expression for current density is obtained as

$$J_v = A'_V e^{-2k_0 d} T e^{-q(\phi_{V0} - \eta V_g - \gamma V_{ds})/k_B T} = A^*_V T e^{-q\phi_V/k_B T} \quad (A8)$$

where $A'_V \propto \frac{16\pi^3 q}{\hbar m^*} g_v k_B \phi_{avg}$, $A^*_V = A'_V e^{-2k_0 d}$, and $\phi_V = \phi_{V0} - \eta V_g - \gamma V_{ds}$. Equation (A8) represents the modified Richardson equation for out of plane electron injection from metal into the 2D semiconductor.

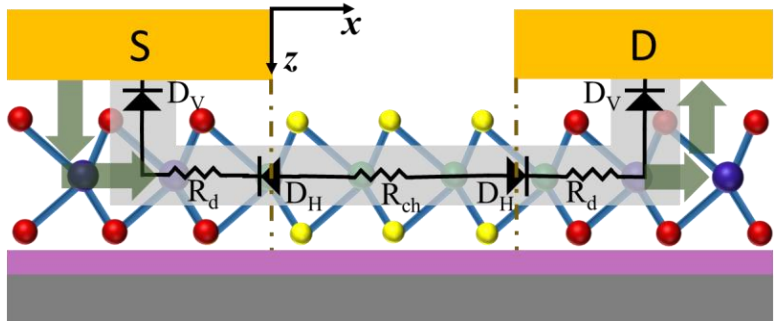

Figure 1. Schematic diagram showing possible carrier injection process from metal to monolayer MoS$_2$. This is modeled as a series connection of back to back diodes and resistor in a back gated device. The arrows indicate direction of electron flow.

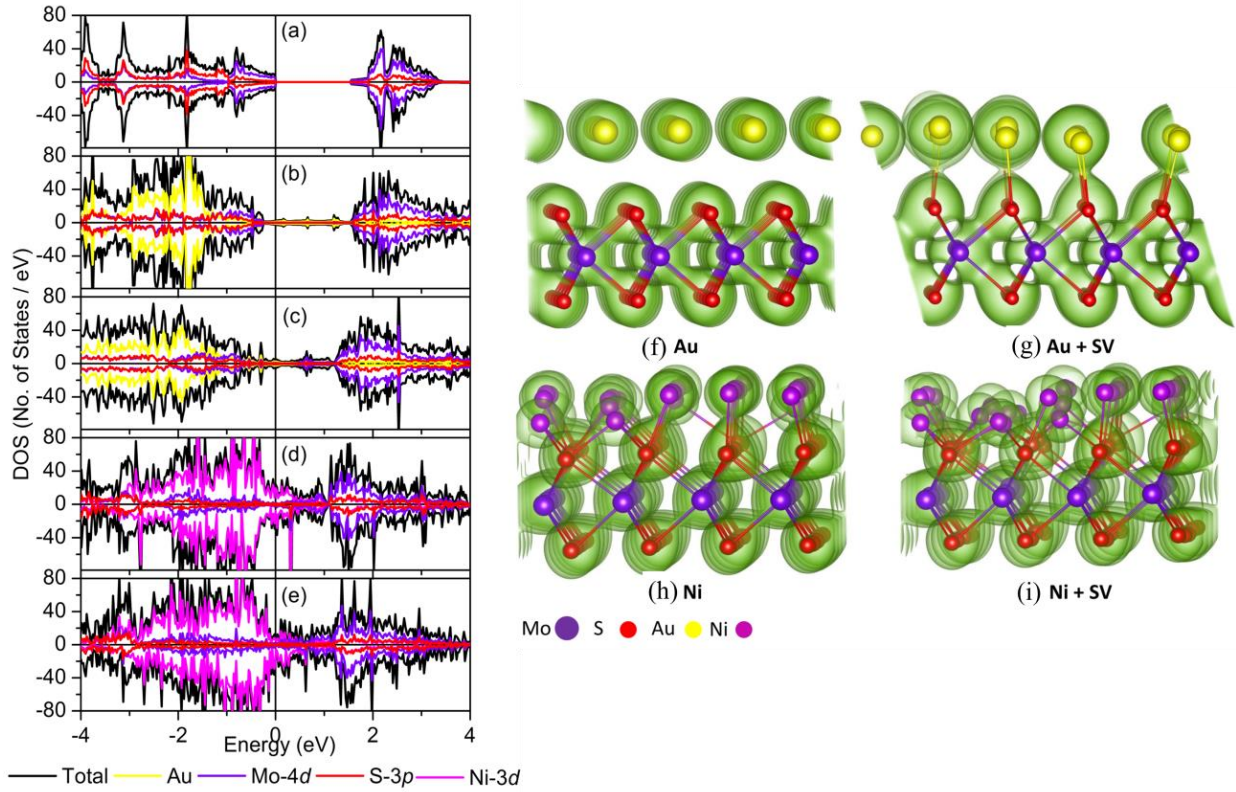

Figure 2. Electronic structure of monolayer $MoS_2$ under monolayer metal. (a) Layer (LPDOS) and orbital (OPDOS) projected density of states for (a) Pristine 1H-$MoS_2$, (b) Au/$MoS_2$ interface, (c) Au/$MoS_2$ interface with interfacial SV, (d) Ni/$MoS_2$ interface, (e) Ni/$MoS_2$ interface with interfacial SV. (f) Charge density plots for Au/$MoS_2$ interface, (g) Au/$MoS_2$ interface with interfacial SV, (h) Ni/$MoS_2$ interface and (i) Ni/ $MoS_2$ interface with interfacial SV.

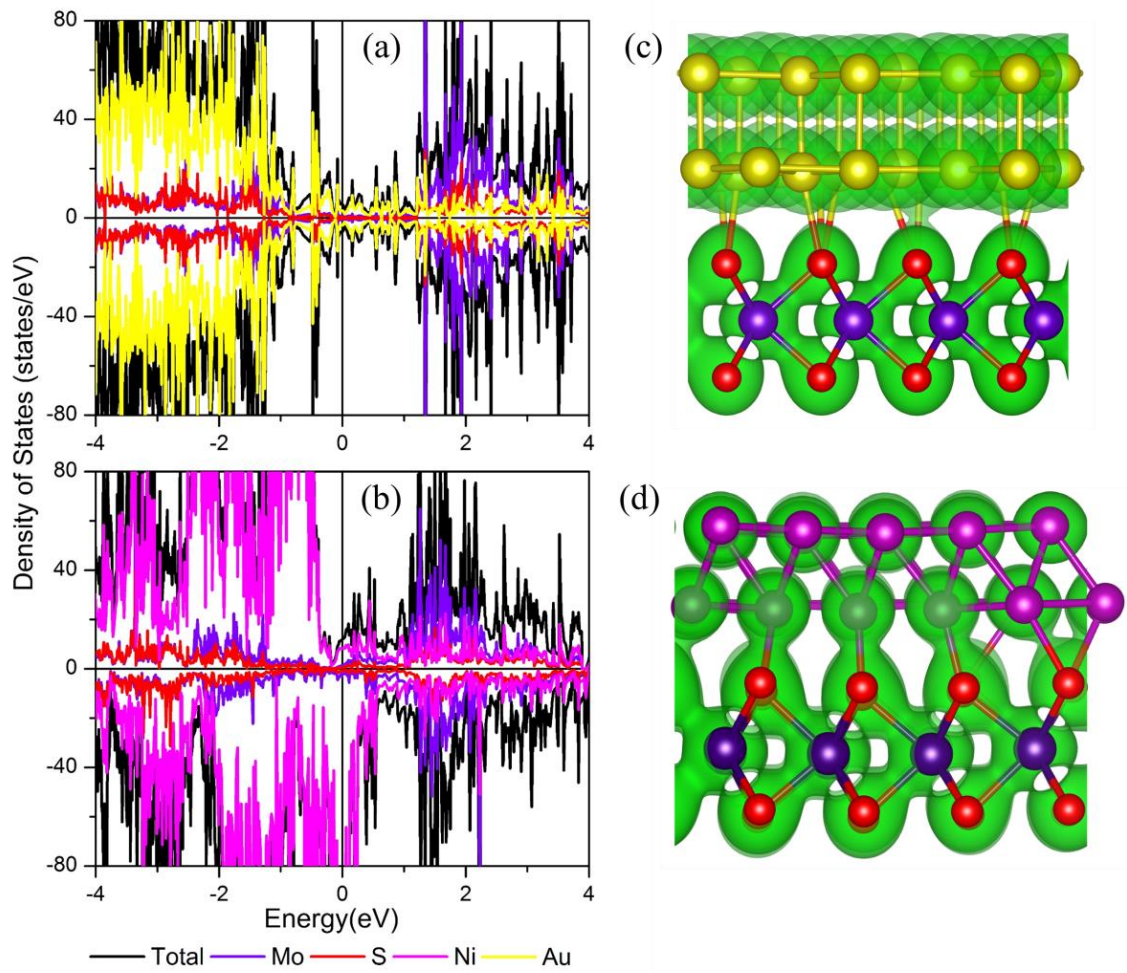

Figure 3. Electronic structure of monolayer MoS$_2$ under bilayer metal. (a) monolayer-MoS$_2$/bilayer-Au DOS, (b) monolayer-MoS$_2$/bilayer-Ni DOS, (c) MoS2/Au charge density plot, (d) MoS2/Ni charge density plot.

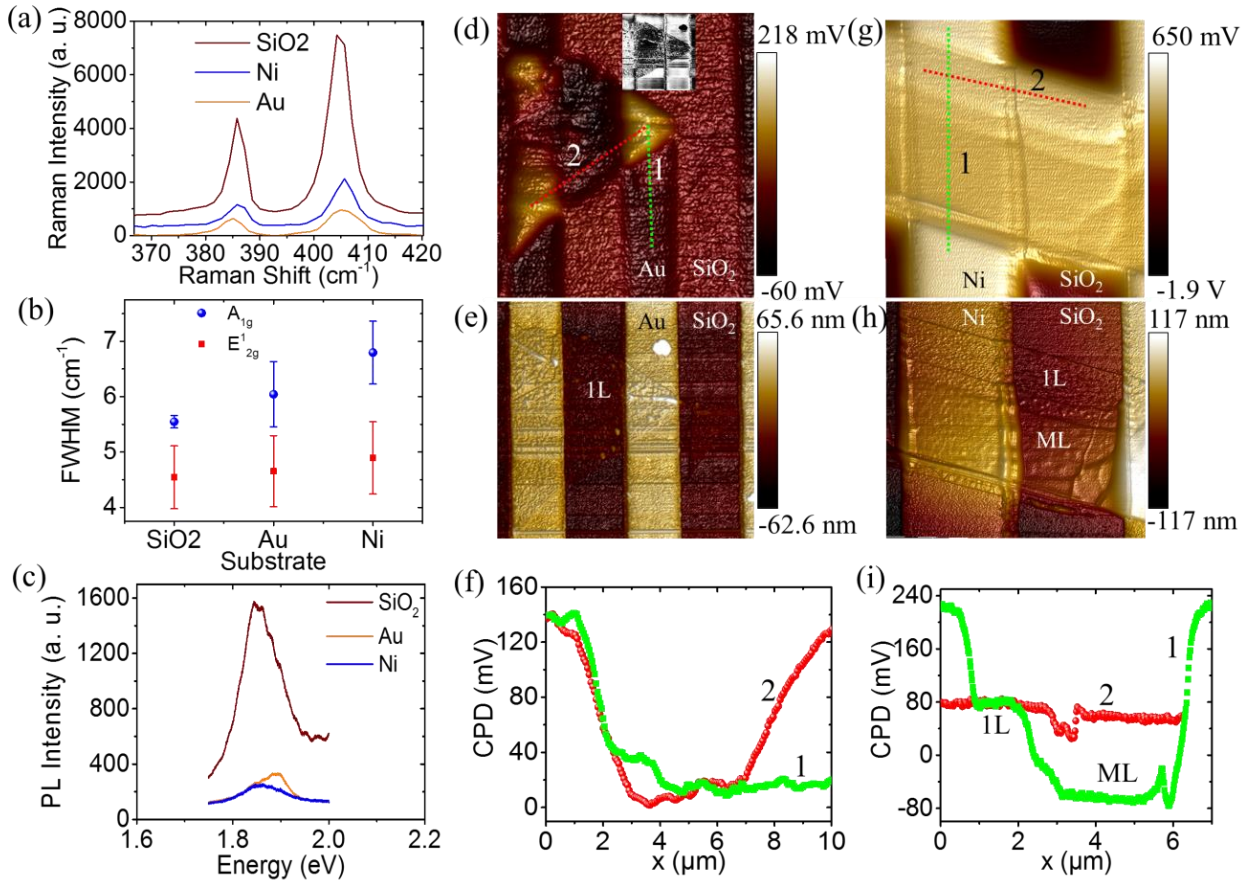

Figure 4. Characterization of monolayer MoS$_2$ on metal and on SiO$_2$. (a) Raman intensity of monolayer MoS$_2$ on various substrates. (b) The FWHM of Raman peaks as a function of substrate. (c) Photoluminescence spectra of monolayer MoS$_2$ as a function of substrate, showing strong suppression on metal. (d) CPD image of monolayer MoS$_2$ on Au and SiO$_2$ lines showing strong work function differences. Inset. A 2D AFM image of the flake. (e) Corresponding AFM image of Au lines and triangular monolayer flake. (f) Values of CPD along scan line 1 (MoS$_2$/Au - Au) and scan line 2 (MoS$_2$/Au – MoS$_2$/ SiO$_2$ – MoS$_2$/Au). (g) CPD image of monolayer/multi-layer MoS$_2$ on Ni and SiO$_2$ lines. (h) Corresponding AFM image of Ni lines and MoS$_2$ flake. (i) Values of CPD along scan line 1 (Ni – 1L MoS$_2$/Ni – ML MoS$_2$/Ni – Ni) and scan line 2 (1L MoS$_2$/Ni – 1L MoS$_2$/ SiO$_2$).

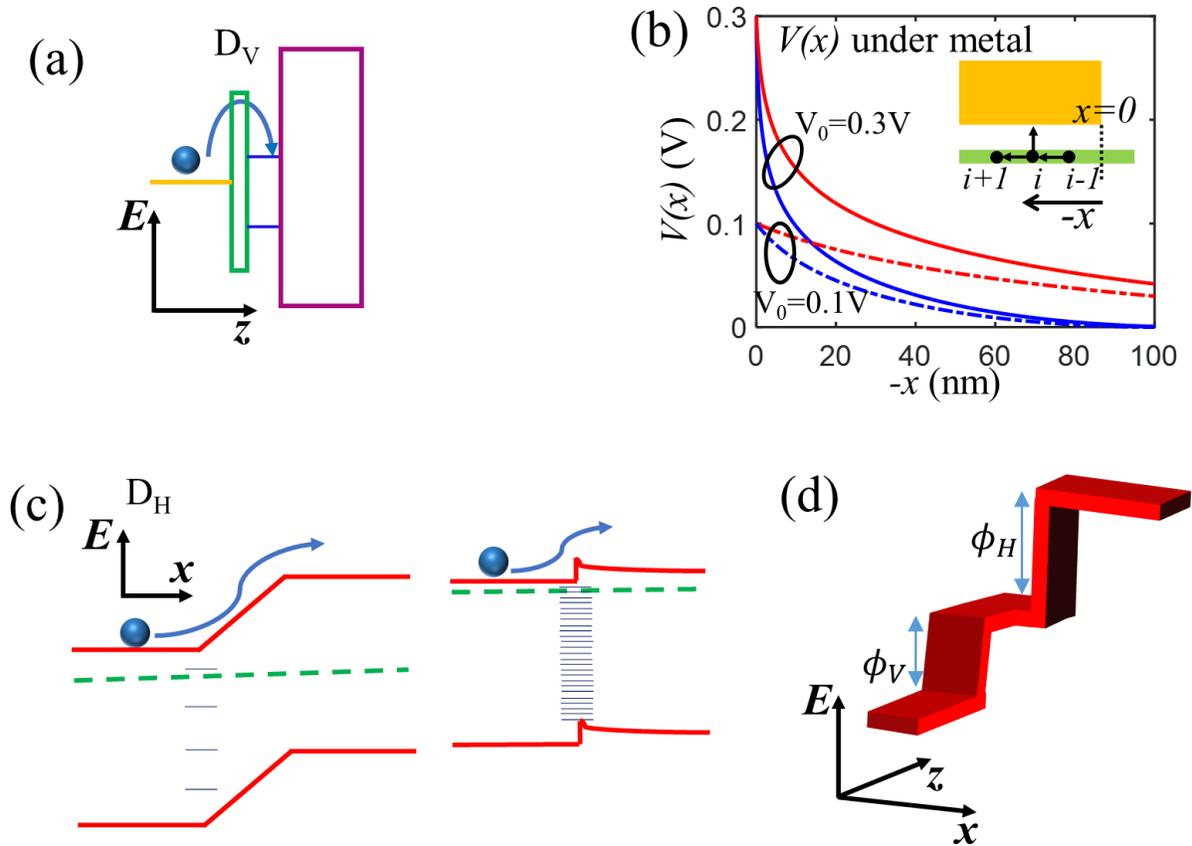

Figure 5. Individual processes in cascade during carrier injection at the source junction. (a) The origin of the vertical diode $D_V$ (with barrier $\phi_V$) between the metal and the monolayer underneath the contact, through a tunneling vdw gap. (b) The variation of potential in the monolayer underneath the contact (current crowding regime) with $V_0 = V(x=0)$. Red curves correspond to 10 times better conductivity in the 2D film, compared with the blue curves. Inset. Schematic of the calculation method in the current crowding regime. (c) Two possible origins of potential barrier $\phi_H$ in horizontal diode $D_H$ at contact edge. Left panel: doping difference induced built-in barrier (weak interaction with contact). Right panel: Fermi level pinning induced Schottky barrier (strong interaction with contact). (d) A schematic three-dimensional diagram depicting the different barriers encountered by the electron injected from the metal at the source end: the two barriers ($\phi_V$ and $\phi_H$) with a sandwiched current crowding region in between.

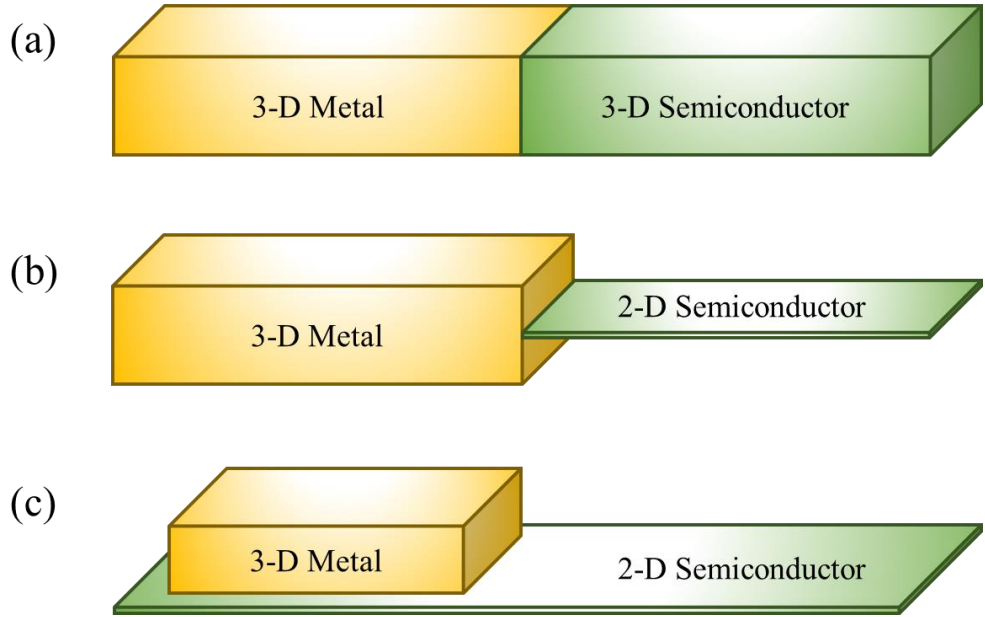

Figure 6. Dimensionality of different metal/semiconductor contact topologies. (a) 3-D metal contacted with 3-D semiconductor. (b) Edge contact between 3-D metal and 2-D semiconductor. (c) Top contact between 3-D metal and 2-D semiconductor.

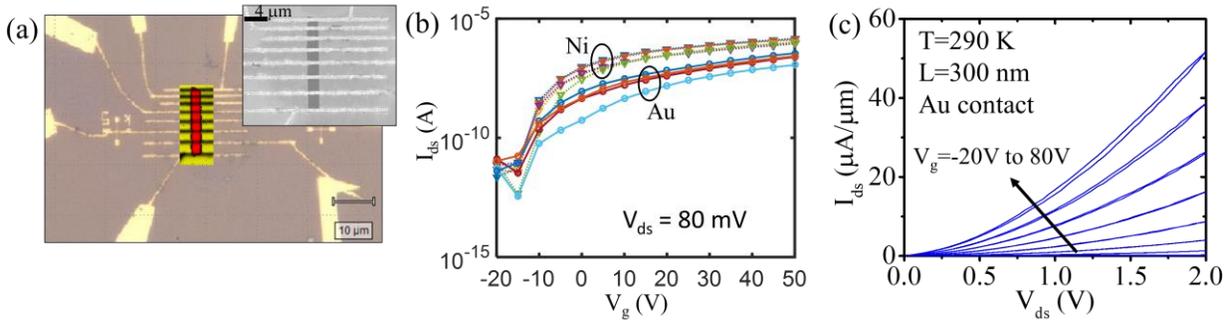

Figure 7. Electrical extraction of effective barrier height. (a) Optical image and the corresponding Raman mapping of a typical TLM device. In the Raman map, red patch corresponds to the monolayer $MoS_2$, and the green signal is from Si. Inset. An SEM image of the device. (b) $I_{ds}$-$V_g$ plot of Au and Ni contacted device, at multiple temperatures. Ni devices show lower threshold voltage. (c) Forward and reverse $I_{ds}$-$V_{ds}$ sweep of a back gated monolayer transistor with 300-nm channel length, showing 52 µA/µm drive current at $V_{ds}$ = 2V and $V_g$ = 80 V.

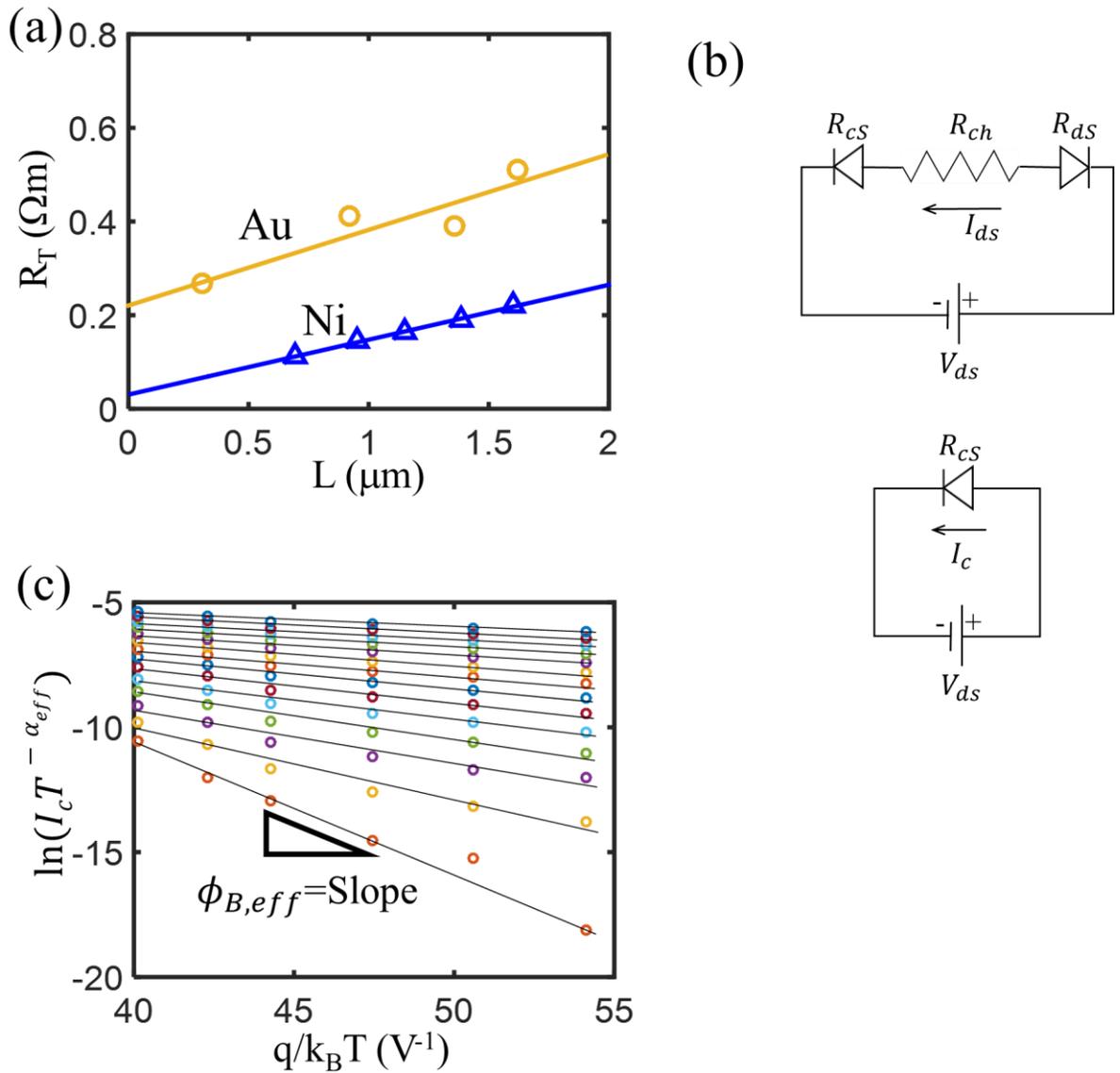

Figure 8. Barrier height extraction methodology. (a) A typical TLM fit for Au and Ni devices with intercept at $L = 0$ corresponding to $R_{cT}$. (b) Method of calculation of $I_c = V_{ds}/R_{cS}$, which is the hypothetical current obtained when $V_{ds}$ is applied across the source contact diode. (c) $I_c$ is used in a Richardson plot at different biasing conditions. The slopes of the linear fits correspond to the effective barrier height.

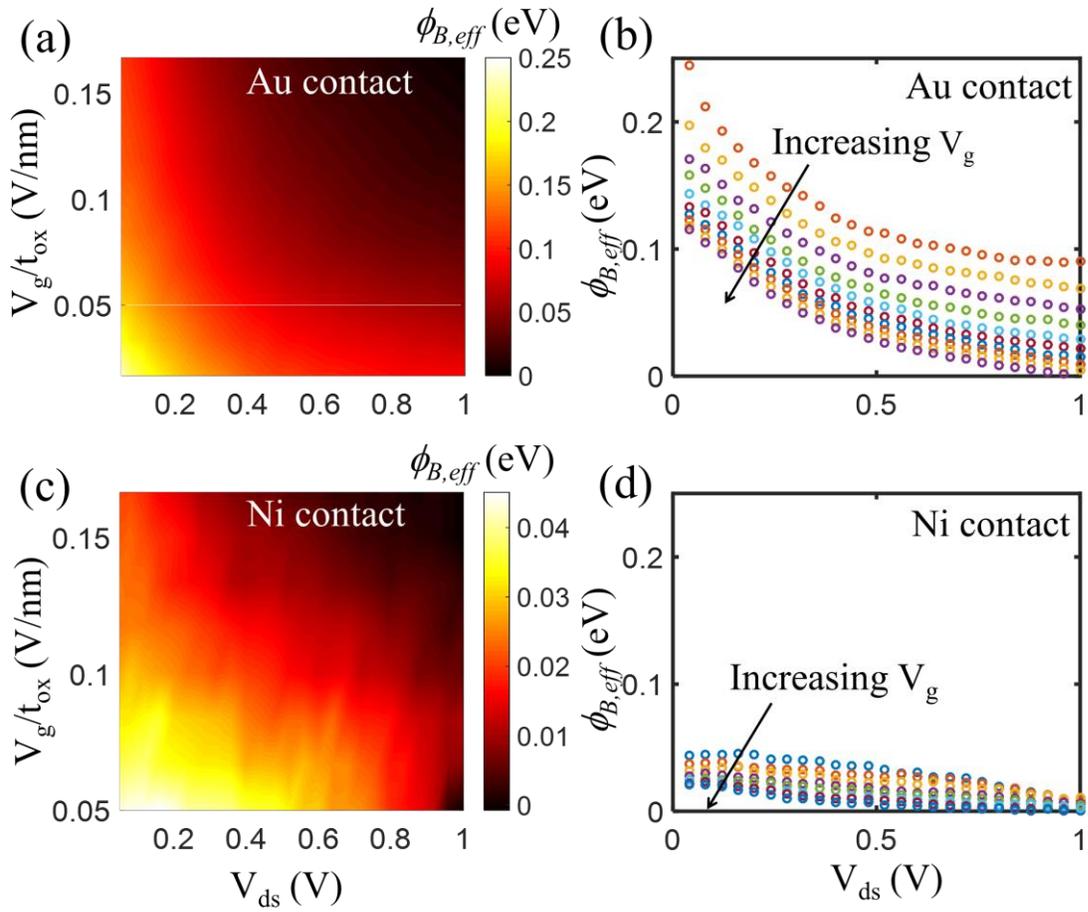

Figure 9. Device operating condition dependent extracted barrier height. (a) The effective barrier height for Au-MoS$_2$ contact, as extracted from TLM, is plotted as a function of $V_{ds}$ and gate electric field. (b) Horizontal slices of Figure (a) at different gate electric field. (c)-(d) The effective barrier height for Ni-MoS$_2$ contact.

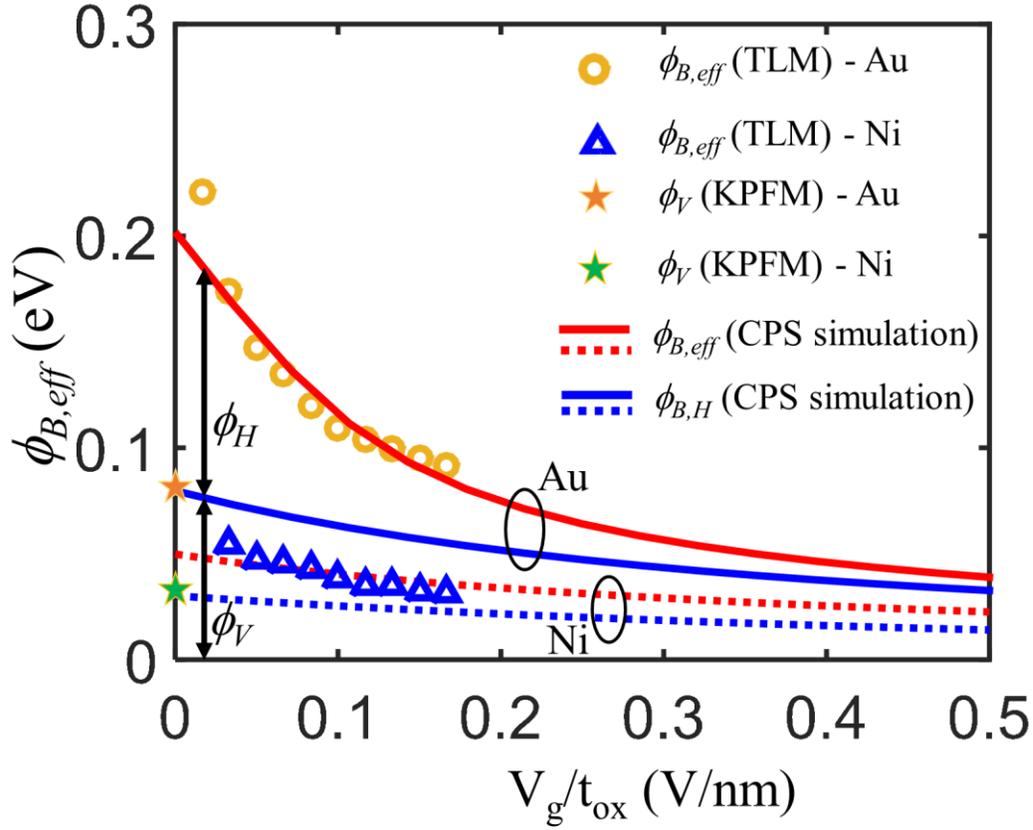

Figure 10. Experimental validation of thermionic model. Comparison of coupled Poisson-Schrodinger (CPS) solution predicted $\phi_{B,eff}$ (shown in red solid line for Au and red broken lines for Ni) and the corresponding TLM extracted value (golden circle for Au and blue triangle for Ni), at small $V_{ds}$ using channel doping ($6.3 \times 10^9$ cm$^{-2}$ for Au and $2.24 \times 10^{12}$ cm$^{-2}$ for Ni) as fitting parameter. The star symbols indicate $\phi_V$ at $V_g = 0$, which is obtained by subtracting the KPFM extracted $\phi_H$ from Simulated $\phi_{B,eff}$. The blue lines (solid for Au and broken for Ni) correspond to $\phi_V$ at non-zero $V_g$, and are obtained by fitting a doping concentration ($7.0 \times 10^{11}$ cm$^{-2}$ for Au and $4.76 \times 10^{12}$ cm$^{-2}$ for Ni) underneath the contact corresponding to the star.

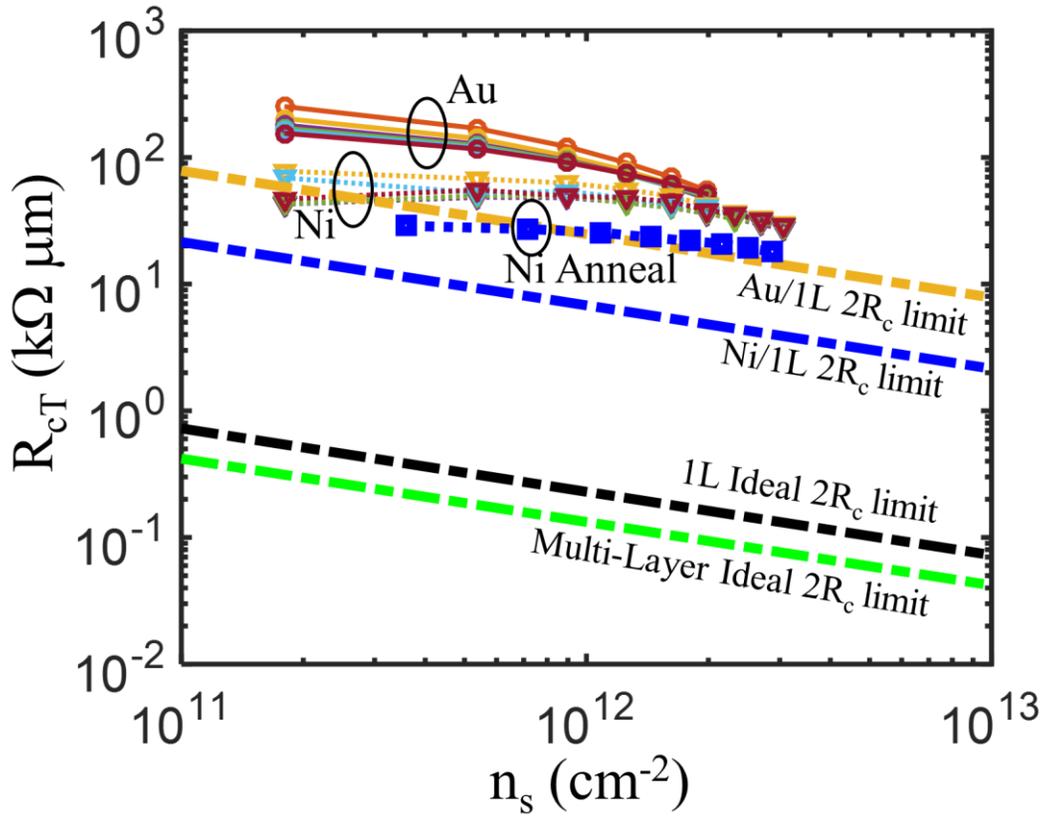

Figure 11. Limits of contact resistance. The extracted total contact resistance $R_{cT} = R_{cS} + R_{cD}$ for Au and Ni at different temperatures (215 K to 290 K) are shown as a function of the 2D sheet carrier density. Annealed Ni contact is also shown at 290 K. The corresponding lower limits achievable for Au ($d = 3$Å, $W_m = 5.3$ eV), Ni ($d = 2.2$Å, $W_m = 5.06$ eV), and an ideal contact ($d = 0$) are shown for comparison. For monolayer (1L) and multi-layer (ML), $g_v = 2$ and $g_v = 6$ are used, respectively. Tunneling vdw gap ($d$) values are obtained from DFT calculation, and metal work function ($W_m$) are obtained from KPFM results.

# Supplemental Material for:

# Nature of carrier injection in metal/2D semiconductor interface and its implications to the limits of contact resistance

*Divya Somvanshi[1,⊥,∥], Sangeeth Kallatt[1,2,∥], Chenniappan Venkatesh[1], Smitha Nair[2], Garima Gupta[1], John Kiran Anthony[3], Debjani Karmakar[4], and Kausik Majumdar[1,*]*

[1]Department of Electrical Communication Engineering, Indian Institute of Science, Bangalore 560012, India

[2]Center for Nano Science and Engineering, Indian Institute of Science, Bangalore 560012, India

[3]Renishaw India, Bangalore 560011, India

[4]Bhabha Atomic Research Center, Homi Bhabha National Institute, Trombay, Mumbai: 400085, India

[⊥]Currently with Department of Physics and Astronomy, Georgia State University, Atlanta, Georgia 30303, USA

[∥]Equal contribution, [*]Corresponding author, email: kausikm@iisc.ac.in

## S1. DFT results for bilayer metal/bilayer MoS₂ system

For the bilayer combined systems, this trend remains the same. The Average distance of metallic layer for Au is ~2.7 A and for Ni is ~2.1 A. With more number of metallic layers, the enhanced electronic effects of metals mask the characters of single MoS₂ layer and its interface with metal. To ensure that the contribution of MoS₂ layer is prominent is DOS figure, we have taken a bilayer MoS₂ system. The bilayer MoS₂/bilayer metal system shows qualitatively similar results as monolayer MoS₂/monolayer metal. Similar to the monolayer metal/monolayer MoS₂ results, the MoS₂ system undergoes an n-type doping. The DOS figure attached below also indicates that with increasing no. of metallic layers, non-magnetic metallic contacts like Au undergoes no qualitative change of interfacial effects. Au -6s electrons are delocalized and getting transferred to the MoS₂-layer and there is almost no spin-polarization within the system. However, for a magnetic metal like Ni, the system becomes more and more spin-polarized with increasing number of metallic layers. The effect of antiferromagnetic Nickel sulphide is getting masked by the spin-polarization of the additional Ni-layer. Therefore, a thicker Ni-layer may be more effective for spin-injection within MoS₂.

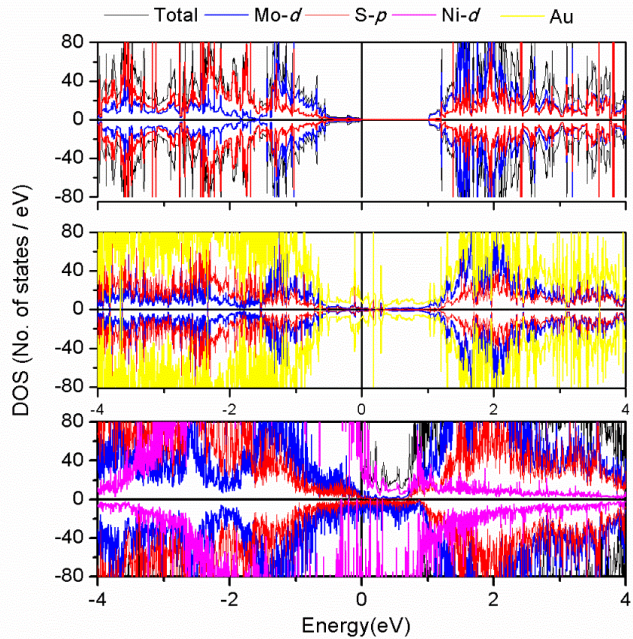

Figure S1. DOS versus energy. Top panel: Bilayer MoS₂, Middle panel: Bilayer MoS₂/Bilayer Au, lower panel: Bilayer MoS₂/bilayer Ni.

## S2. XPS analysis of MoS$_2$ flakes on metal and on SiO$_2$ and comparison with DFT results

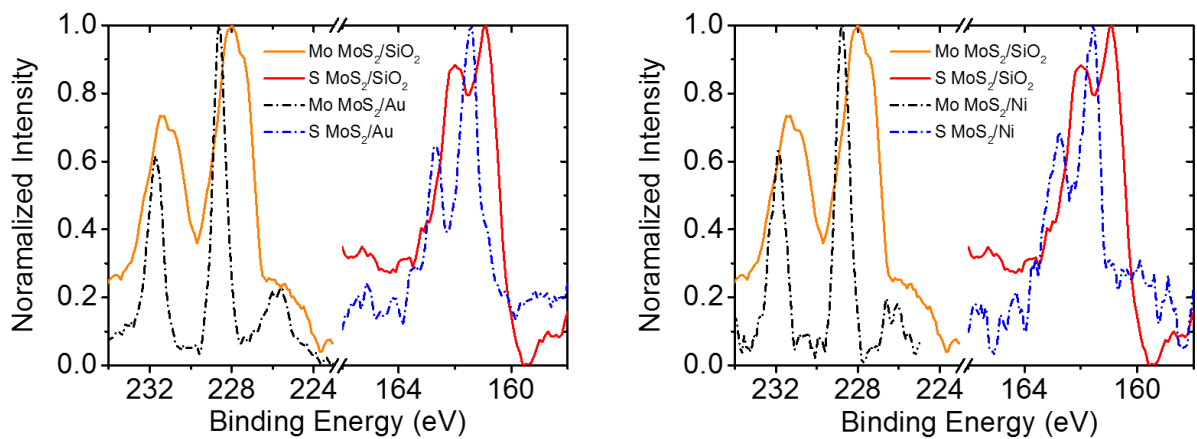

Figure S2. XPS analysis shows the binding energies of the Molybdenum and Sulfur core levels are blue shifted for MoS$_2$ samples on metals (Au and Ni), compared with the MoS$_2$ samples on SiO$_2$. The amount of shift is ~ 0.7 – 0.8 eV. Such a shift indicates that the Fermi level (which is the reference for the measurement) is relatively shifted upward, closer to the conduction band, and hence implies that the samples on metals are doped more *n*-type compared with samples on SiO$_2$.

**Estimated upward shift of Fermi level from DFT calculation**:

| System | Relative shift of $E_F$ (eV) |
|---|---|
| Au/ MoS$_2$ | +0.86 |
| Au/ MoS$_2$ + SV | +2.33 |
| Ni/ MoS$_2$ | +1.96 |
| Ni/ MoS$_2$ + SV | +2.08 |

Table S2: Relative $E_F$ shift for different interfaces with respect to the pristine 1H- MoS$_2$ as evaluated from the DFT calculation.

## S3. Estimation of image force induced barrier lowering in monolayer 2D material

Convolution of source and the green's function gives the potential due to any arbitrary charge distribution when solved with the appropriate boundary conditions. The results for the three dielectric configuration shown in Figure S3-1. helps to understand the role of image forces in altering the potential due to the point charge.

The image force terms arise from the difference in the dielectric surroundings which modifies the potential profile. The voltage in the middle region has the form [1]:

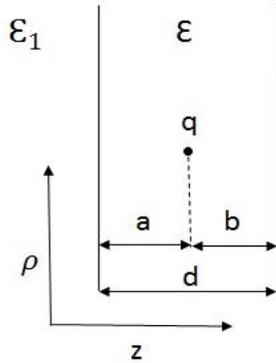

$$V = \frac{q}{4\pi\epsilon}\left[\frac{1}{(\rho^2+z^2)^{1/2}} + \sum_{n=0}^{\infty}(k_1 k_2)^n \left(\frac{k_1}{[\rho^2+(z-2a-2nd)^2]^{1/2}} + \frac{k_2}{[\rho^2+(z-2b-2nd)^2]^{1/2}} + \frac{2k_1 k_2}{[\rho^2+(z-2a-2b-2nd)^2]^{1/2}}\right)\right],$$

Here $k_1$ and $k_2$ are the reflection coefficients at the 2 interfaces defined as:

$$k_1 = \frac{(\epsilon-\epsilon_1)}{(\epsilon+\epsilon_1)} \text{ and } k_2 = \frac{(\epsilon-\epsilon_2)}{(\epsilon+\epsilon_2)}.$$

Figure S3-1

The terms under the summation in the above potential expression are the contributions due to the charges induced at the two interfaces, i.e. due to the image force of the point charge, the nature of polarization being determined by the sign of the reflection coefficients.

The potential energy is then obtained by evaluating the integral: $E = \int_0^q V \, dq$

The change in energy at the location of the point charge obtained by putting taking $\rho = z = 0$ is:

$$E = \frac{q^2}{16\pi\epsilon}\sum_{n=0}^{\infty}(k_1 k_2)^n \left(\frac{k_1}{a+nd} + \frac{k_2}{b+nd} + \frac{2k_1 k_2}{a+b+nd}\right)$$

The image force terms will either strengthen or reduce the potential profile depending on the signs of $k_1$ and $k_2$, and i.e., depending on whether the dielectric constant value of the surrounding media is lower or higher than the middle region.

In our particular case of monolayer $MoS_2$ portion underneath the contact metal, replacing medium 1 by a conductor and if $\epsilon \sim \epsilon_2$ [approximately holding true for $MoS_2$ (4.0) and $SiO_2$ (3.9)], then the energy change is further simplified by putting $k_1 = -1$ and $k_2 \approx 0$ (Figure S3-2):

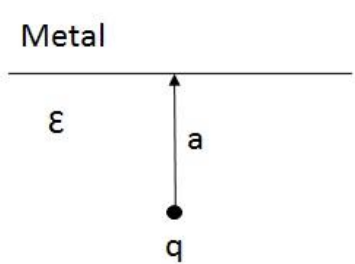

$$E = \frac{-q^2}{16\pi\epsilon a}$$

Here, we have assumed the van der Waals gap has a dielectric constant which is average of $MoS_2$ and the other medium. The final expression yields about 0.25 eV suppression of Schottky barrier height using $a = 3.5$Å.

Figure S3-2

## S4. Extraction of barrier height with $\alpha_{eff} = 1.5$

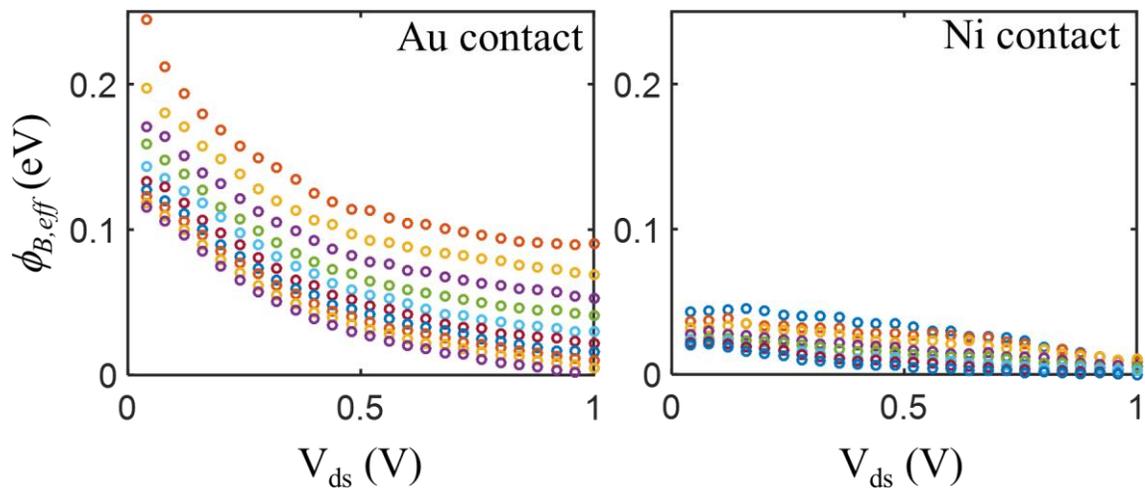

Figure S4. Extracted barrier height for Au and Ni contacts with monolayer MoS$_2$ as a function of V$_{ds}$ for different V$_g$.

## S5. 1D Coupled Poisson-Schrodinger solution to find barrier height

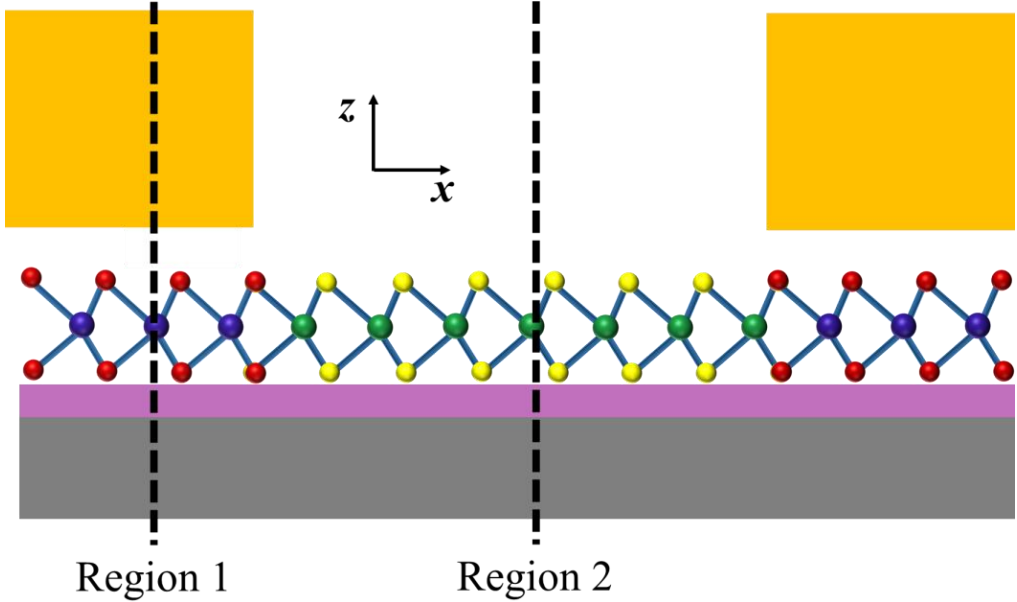

Figure S5. Schematic diagram showing the two regions (underneath the contact and in the channel) where the simulations are performed.

For a potential distribution $V(z)$ along the vertical direction, the 1D Schrodinger equation can be solved to obtain the energy eigenvalues $E_i$ and the corresponding wave functions $\psi_i(z)$ as:

$$\left( \frac{\hbar^2}{2m^*} \frac{d^2 \psi(z)}{dz^2} + qV(z) + E \right) \psi(z) = 0$$

The electron volume density profile $n(z)$ can now be obtained by summing over all sub-bands $i$ as

$$n(z) = \sum_i \int_{E_i}^{\infty} d\varepsilon \, f(\varepsilon) D(\varepsilon) |\psi_i(z)|^2$$

where $f(\varepsilon) = \frac{1}{1+e^{[\varepsilon-\mu]/k_B T}}$ and $D(E) = g_v \frac{m^*}{\pi \hbar^2}$. Similarly, the total hole density distribution is also obtained. The Poisson equation is now solved to calculate the new potential:

$$\frac{d^2 V(z)}{dz^2} = \frac{q[n(z) - p(z) + N_a - N_d]}{\epsilon_0 \epsilon_r}$$

Appropriate boundary conditions are used at the back gate and at the other side (metal contact or air, depending on region 1 or region 2 in Figure 1). $\epsilon_r$ is assumed to be 4. For our simulation, we assumed $N_a = 0$, and use $N_d$ as a fitting parameter to match the experimental data. In particular, two different $N_d$ are used for simulating the barrier height underneath the metal contact and in the channel region. Once the voltage at the two portions is known, the effective barrier height can be predicted directly as the difference between the conduction band edge and Fermi level position.

## S6. Contact resistance from a different set of TLMs

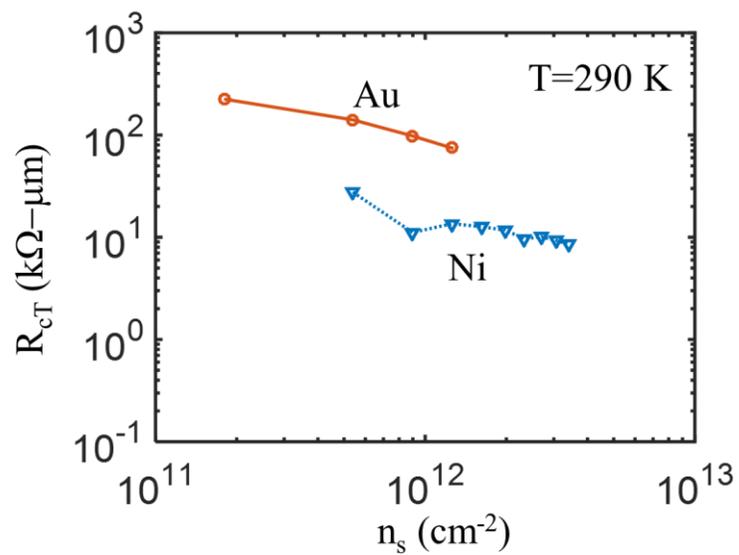

Figure S6. Measured total contact resistance (source side plus drain side) as a function of 2D sheet carrier density at T=290 K, obtained from another set of Au TLM and Ni TLM.

## S7. Contact resistance dependence on $V_{ds}$

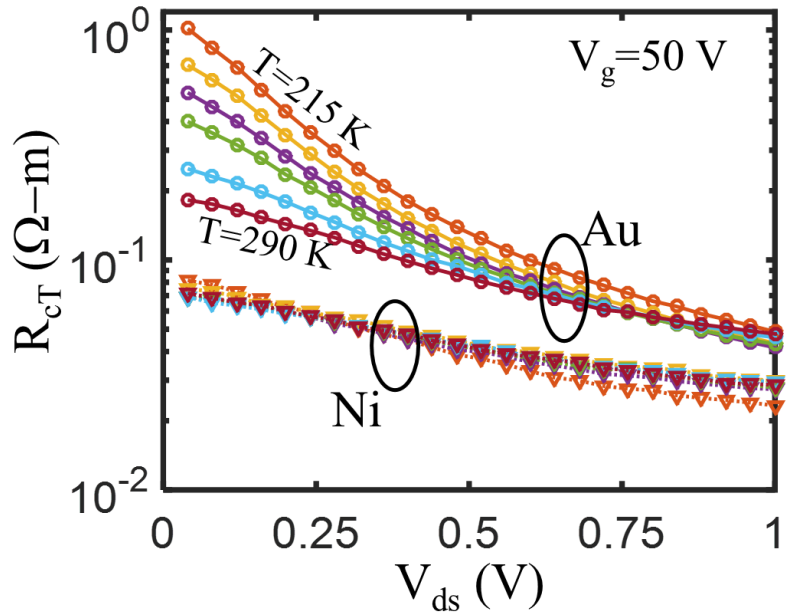

Figure S7. The extracted total contact resistance (source side plus drain side) is plotted as a function of drain bias, for different measurement temperatures. Reverse temperature trend is observed at smaller and larger $V_{ds}$, both for Au and Ni contacts, providing insights to the different dominant mechanisms (thermionic emission efficiency versus scattering of electrons) controlling the contact resistance at the two regimes.